# Population density and vegetation resources influence demography in a hibernating herbivorous mammal

Anouch TAMIAN[1], Vincent A VIBLANC[1]†, F Stephen DOBSON[1,2] and Claire SARAUX[1*]†

[1]*Institut Pluridisciplinaire Hubert Curien, CNRS, 23 Rue du Loess, 67037 Strasbourg, France*

[2]*Department of Biological Sciences, Auburn University, Auburn, Alabama, USA*

*Corresponding author: Claire Saraux, Université de Strasbourg, Centre National de la Recherche Scientifique, Institut Pluridisciplinaire Hubert Curien, Unité Mixte de Recherche 7178, 67087 Strasbourg, France. claire.saraux@iphc.cnrs.fr

†Claire Saraux and Vincent A Viblanc contributed equally to the work reported here.

*Author Contributions*

AT, CS and VAV designed the study. AT, CS, VAV, and FSD collected the data. AT and CS analyzed the data. AT wrote the manuscript. All authors contributed critically to the drafts and gave final approval for publication.

**ABSTRACT**

Demography of herbivorous mammal populations may be affected by changes in predation, population density, harvesting, and climate. Whereas numerous studies have focused on the effect of single environmental variables on individual demographic processes, attempts to integrate the consequences of several environmental variables on numerous functional traits and demographic rates are rare. Over a 32-year period, we examined how forage availability (vegetation assessed through NDVI) and population density affected the functional traits and demographic rates of a population of Columbian ground squirrels (*Urocitellus columbianus*), an herbivorous hibernating rodent. We focused on mean population phenology, body mass, breeding success and survival. We found a negative effect of population density on demographic rates, including on breeding success and pup and adult survival to the next year. We found diverging effects of vegetation phenology on demographic rates: positive effects of earlier start to growing season on adult female and juvenile survival, but no clear effect on male survival. Interestingly, neither population density nor vegetation affected population phenology or body condition in the following year. Vegetative growth rate had a positive influence on female mass gain (somatic investment) over a season, but vegetative growth rate and biomass, surprisingly, had negative effects on the survival of young through their first hibernation. Later vegetative timing during the year had a positive influence on survival for all ground squirrels. Thus, ground squirrels appeared to benefit more from later timing of vegetation than increases in vegetative biomass per se. Our study provides evidence for complex ecological effects of vegetation and population density on functional traits and demographic rates of small mammal populations.

**INTRODUCTION**

Understanding the mechanisms that influence animal life history and demography is a central focus in ecology. These mechanisms have been especially well studied in rodents and other mammals (Blair 1953; Erb et al. 2001). They include competition for resources (e.g., forage such as herbage), avoiding predation, parasites, and diseases, and coping with changes in the physical environment, such as seasons and climate. Such processes are multifactorial in nature (Wallner 1987; Abbiati et al. 1992). For instance, variation in competition for food involves changes in the number of competitors (population density), and in the amount, availability and quality of food resources (Putman et al. 1996; Solberg et al. 2001; Lima et al. 2003; Saether et al. 2004; Le Galliard et al. 2010).

Forage availability can have strong effects on consumer body mass, survival, and reproductive timing and success (also known as the "bottom-up effect", Gruner et al. 2008, Preisser 2008). For primary consumers (herbivores) vegetation biomass, quality, rate of growth, and phenology (the timing of vegetation growth) should be important factors affecting functional traits and demographic rates, especially in seasonal environments where few food resources are available in winter (Hansson 1971; O'Connell 1989; Jędrzejewski and Jędrzejewska 1996; Korslund and Steen 2006). In such environments, life cycles have often evolved so that reproduction coincides with brief periods of high forage availability (Williams et al. 2017a; 2017b; Rehnus et al. 2020). Species that use daily foraging during such productive periods to support reproduction are termed "income breeders" (Jönsson 1997; Broussard et al. 2005).

The dynamics of vegetation growth are important aspects during these periods. Vegetation growth, quality, and timing are strongly influenced by climate (Voigt et al. 2003; Rosenblatt and Schmitz 2016). Vegetation growth is usually higher with warmer temperatures and increased precipitation (He et al. 2012; Verbyla and Kurkowski 2019), making it susceptible both to interannual weather variability and climatic changes. Further, with global warming, a general advancement of plant phenology has been observed in a variety of ecosystems (Cleland et al. 2007; Piao et al. 2019). Non-concomitant changes in the phenology of different trophic levels can lead to mismatches between periods of optimal food availability and the timing of reproduction, negatively affecting individual reproductive success and survival, with consequences for population size (Plard et al. 2014) and viability (Cohen et al. 2018; Radchuk et al. 2019; Schano et al. 2021). Changes are particularly marked in Arctic and alpine ecosystems, where climates are rapidly changing (Walther et al. 2005; Thackeray et al. 2010; Finch 2012; Gottfried et al. 2012; Zheng et al. 2022). This is critical for hibernating species inhabiting those regions, since reproduction and preparation for subsequent hibernation depend on the phenology, amount and quality of vegetation growth during the temporally restricted food bursts during their active season (Tchabovsky et al. 2005).

Besides the influences of food production (timing and quantity), the access to resources in a given area often depends on population density (Clutton-Brock and Harvey 1978). Intra-specific competition for food resources is a process that may compound the negative effects of phenological mismatches between forage availability and reproduction, depending on yearly changes in population density (Bomford and Redhead 1987; Reed et al. 2015; Ross et al. 2018). Population density has also been widely shown to affect reproduction and survival, including through effects on individual stress and body condition (Arcese and Smith 1988; Eccard and Ylönen 2001; Bonenfant et al. 2002; Creel et al. 2013; Tveraa et al. 2013; Reed et al. 2015, but see Boonstra and Boag 1992). However, to date, comprehensive studies investigating the effects of forage availability and population density on several functional traits and demographic rates simultaneously are scarce (Violle et al. 2007; Gamelon et al. 2017).

We studied temporal trends in functional traits and demographic rates of a population of a small hibernating rodent, the Columbian ground squirrel (*Urocitellus columbianus*), over a 32-year period. We investigated the hypothesis that changes in forage availability and accessibility affected functional traits and demographic rates. Specifically, we tested how resource availability (estimated using the Normalized Difference Vegetation Index, NDVI; Pettorelli et al. 2005, 2011; Hurley et al. 2014; Rézouki et al. 2016) and population density influenced juvenile and adult survival, breeding success, phenology, and body condition of individuals in our population. To do so, we (1) described changes in population density, (2) described yearly fluctuations in vegetation phenology and biomass, (3) described temporal trends in functional traits and demographic rates, and (4) evaluated the contributions of forage availability and population density to functional traits and demographic rates.

Columbian ground squirrels are hibernating, colonial rodents that inhabit open meadows of the Rocky Mountains (Elliott and Flinders 1991). Their active period is notably short, as they hibernate from August to April of the following year, about 70% of the year (Dobson et al. 1992). These generalist herbivores consume a variety of grasses and forbs, including clover (*Trifolium dubium*), dandelion (*Taraxacum officinale*), bentgrass (*Agrostis alba*) and Kentucky bluegrass (*Poa pratensis*) (Harestad 1986; Ritchie 1988). Foraging is especially important during the spring and summer activity period, when reproduction and fattening for subsequent hibernation occur (Young 1990). Ground squirrels spend the vast majority of their time above ground feeding during the active season (Ritchie 1990; Tamian et al. 2023; Tamian et al, unpublished data). Experimental studies have shown that food supplementation during the active period increases juvenile survival, the size of litters produced, and adult body mass (Dobson and Kjelgaard 1985a, b; Dobson et al. 1986; Dobson and Murie 1987; Dobson 1988). Moreover, seasonal climate influences female annual fitness, a positive association being found between annual fitness, and spring temperature and early summer rainfall (Lane et al. 2012;

Dobson et al. 2016). The underlying assumption is that such climate conditions should provide favorable conditions for grass and herb growth, as suggested by subsequent positive impacts on reproduction and survival (Ritchie 1988, 1990). Similarly, climate conditions during the active season affect ground squirrels as they enter hibernation, by modifying the extent of reserves that are stored as fats before hibernation (Zammuto and Millar 1985).

The demographics and size of Columbian ground squirrel populations are primarily regulated by food availability (Dobson and Kjelgaard 1985a; Dobson, 1995; Dobson & Oli, 2001). Thus, relaxing competition for food in years of low population density was expected to increase individual success, through higher mass gain over the season and consequently better survival, as well as higher reproductive success. These expectations were tested by comparing population density to the demographic rates and functional traits. In addition, food resource availability was expected to have direct effects on individual condition. Active seasons with early and high vegetation growth and biomass that are timed with the reproductive and hibernating phenology of ground squirrels were expected to lead to higher body mass, reproductive success, and survival for this primarily income breeder (Dobson et al. 1999; Broussard et al. 2005). We used NDVI to calculate vegetation indices reflecting the timing and rate of spring greening-up, as well as yearly overall vegetation biomass available to generalist herbivores like the ground squirrels, to disentangle these processes. Our expectations about the effects of vegetation were tested by comparing indices based on NDVI to the demographic rates and functional traits.

Carry-over effects from a current active period to the subsequent annual active period were also expected. A low density population with less competition, associated with high forage growth and biomass, was expected to allow individuals to store more body reserves for hibernation, positively affecting individual mass the next spring (Murie and Boag 1984; Dobson et al. 1992; Rubach et al. 2016). We tested for these carry-over effects by examining whether current-year conditions appeared to influence subsequent-year functional traits of the ground squirrels. Such a mechanism might also influence how individuals adjust their emergence timing to yearly changes in additional environmental conditions (e.g., timing of snowmelt, Lane et al. 2012; Dobson et al. 2016; Tamian et al. 2022; Thompson et al. 2023).

**MATERIALS AND METHODS**

*Long-term population monitoring*

Columbian ground squirrels were monitored from 1992 to 2023 in the Sheep River Provincial Park, Alberta, Canada. The study site is a southeast-facing 2.6-ha meadow in the foothills of the Rocky Mountains (50°38'N, 114°39'W; 1500-1540 m elevation), surrounded by mixed forests (primarily lodgepole pine, white spruce, quaking aspen, and balsam poplar), and composed of vegetation representative of grasslands in the montane sub-alpine (Alberta Parks 2008). Ground squirrel are generalist foragers, their diets mainly composed of several species of grasses and forbs (Harestad 1986; Ritchie 1988, 1990; Elliott and Flinders 1991). Individuals were captured using live-traps (National Live Traps Tomahawk Co., Hazelhurst, WI, USA: 13 x 13 x 40 cm³), baited with a small amount of peanut butter. Ground squirrels were first captured as juveniles at about the time of weaning, or when they first appeared on the meadow as immigrant adults. All ground squirrels were permanently marked with unique numbered metal ear tags (Model no. 1, National Band & Tag Co., Newport, KY). Sex was determined by visual inspection of genitalia (Murie and Harris 1982).

Each year, the population was monitored daily from before the first emergences from hibernation (mid-April to early May) to the end of lactation (mid-July). In spring, the first day of observation or capture (usually the same) for each ground squirrel was used to estimate the hibernation emergence date, typically confirmed by the squirrel's appearance and physical condition (presence of large skin flakes and abdominal skin flaps where fat reserves had been lost, no defecation upon capture; Murie and Harris 1982). Emergence date was recorded as ordinal (number of days after the 1st of January each year), and each ground squirrel was weighed (±5g) using a Pesola® spring-slide scale at initial capture (emergence mass, EM), and given a unique individual dye mark (Clairol® Hydrience N°52 Black Pearl, Clairol Inc., New York, USA) on its dorsal pelage for later visual identification in the field. Reproductive females were monitored until the first emergence of their juvenile offspring from natal nest burrows, around the time of weaning. At that time, females were weighed (±5g), and all pups were captured, sexed, weighed (±1g), and ear-tagged. Litters were subsequently observed over 1-3 days to ensure that all pups had been captured.

The primary emigrants from the population were yearling males, and there was little recorded female immigration or emigration (Wiggett et al. 1989; Wiggett and Boag 1989; Festa-Bianchet and King 2011). Because the entire population was trapped each year (confirmed by extensive visual observations from 3-m-high benches and daily trapping until only dye-marked individuals remained in the spring population), we were able to determine the survival of adults, juveniles and yearling females (but not yearling males, due to emigration) from one year (active period) to the next.

*Population density and growth*

Each year, from 1992 to 2023, we counted all individuals in the population that emerged from hibernation, including male and female adults and yearlings, but excluding offspring born later within the year (pre-breeding census). Because the meadow was delimited to a fixed surface of ca. 2.6 ha, 'population density' has a consistent relationship to population size and both terms are used interchangeably in our study. Further, as the entire population was trapped each year (see above), no associated error had to be considered with the counts. The finite rate of population growth was empirically assessed as the ratio of spring population density in year t+1 divided by spring population density in year t.

*Ethic statement*

Research followed ASM guidelines, authorizations for conducting research and collecting samples in Sheep River Provincial Park were obtained from Alberta Environment and Parks and Alberta Tourism, Parks, and Recreation. Animal care was carried out in accordance with Auburn University IACUC protocol, with additional approval from the University of Calgary and Alberta Fish & Wildlife.

*Data analyses and statistics*

All data analyses and statistics were performed in R v.4.1.2 (2021-11-01). Results are presented as means ± SE, along with the number of observations (n). Where appropriate, we ensured model residuals were normally distributed by visual inspection of density distributions, Q-Q plots, cumulative distribution functions, and P-P plots using the "fitdistrplus" package in R (Delignette-Muller and Dutang 2015).

*Yearly vegetation growth and biomass*

To quantify yearly vegetation growth and biomass and examine its effects on ground squirrel demography and life history traits, we used the Normalized Difference Vegetation

Index (NDVI) (Pettorelli et al. 2005, 2011; Hurley et al. 2014; Rézouki et al. 2016). This index is particularly appropriate for generalist herbivores like Columbian ground squirrels. NDVI is derived from the difference between the visible red (RED) and near-infrared (NIR) light reflectance of satellite images. It is calculated as NDVI = (NID-RED) / (NID+RED) and ranges from -1 to 1. NDVI assesses the amount (biomass) of live green vegetation from satellite images: chlorophyll from vegetation absorbs RED whereas mesophyll in the leaves reflects NIR (low values of NDVI correspond to absence or limited vegetation).

Daily NDVI data for the study site (Lat: 50°38'N, Long: 114°39'W) were downloaded from the Administration-Advanced Very High Resolution Radiometer (AVHRR, up to 2013) and Visible Infrared Imaging Radiometer Suite (VIIRS, from 2014) datasets of the U.S. National Oceanic and Atmospheric Administration (Vermote 2019) for the period spanning 1991-2023 (https://www.ncei.noaa.gov/data/land-normalized-difference-vegetation-index/access/). NDVI as a proxy for vegetation growth on our study site was validated against empirical measures obtained in 2022 and 2023 (see Supplementary Data SD1: Fig S1 and S2). In each year, vegetation quality and quantity (overall productivity and biomass), and vegetation phenology including timing of greening and the rate of vegetation growth were indexed from NDVI (Pettorelli et al. 2005). Because some daily NDVI data were missing from the data set (possibly due to cloud cover), we averaged NDVI by week.

We then calculated (1) mean yearly NDVI by averaging weekly values during the active period of the animals (from the first emergences from hibernation: week 14, to the approximate period of immergence: week 34), (2) NDVI rate of increase within each year and (3) an index of the seasonal phenology (starting week of the increase period of NDVI). The mean yearly NDVI provided us with a measure of yearly vegetation production and overall biomass during the squirrels' active season, whereas the rate and dates of NDVI increase within each year allowed us to investigate vegetation phenology by determining the rate of spring greening-up and the yearly onset of NDVI increase to identify early vs. late years (Pettorelli et al. 2005).

Details on the methodology used to estimate NDVI phenology and growth speed can be found in Supplementary Data SD1: Fig. S3. Briefly, we first smoothed the weekly data using splines. Then, we identified the period of highest increase in NDVI through breakpoint analyses of the derivative of this smooth with an added constraint of NDVI > 0.1, as a NDVI below 0.1 is usually considered as bare of vegetation (rocks, snow, etc.). The period of highest derivative (and NDVI above 0.1) was considered as the vegetation-growing period. The start date of this period was considered as the index of phenology, and the mean derivative over this period was considered the growing slope of the year. The phenology index might occur before ground squirrels emerged from hibernation because NDVI reflects the overall greening of vegetation (grasses and forbs, but also trees) over a relatively large area (0.05°x0.05° grid, i.e. 5*3km in our study area). We used NDVI as an overall index of vegetation to characterize interannual variations in green-up phenology, as has been applied previously (Pettorelli et al. 2005).

*Temporal changes in population demographic rates and functional traits*

We examined how different demographic rates and functional traits varied over time. For demographic rates, we considered rates pertaining to reproduction (including mean annual litter size at weaning, mean pup survival of the offspring from weaning through their first hibernation, proportion of reproductive females in the population), and survival (including mean annual adult (2+) male and female survival, and yearling female survival). Survival could not be determined accurately for yearling males due to dispersal (Wiggett et al. 1989; Wiggett and Boag 1989; Festa-Bianchet and King 2011). The proportion of reproductive females in the population was calculated as the annual percentage of adult mated females that weaned a litter.

For functional traits, we considered traits related to phenology (mean adult male and female dates of emergence from hibernation), or body condition averaged annually (mean adult male and female body mass at emergence from hibernation, and mean adult female mass gain between emergence and weaning of offspring).

We tested patterns of change in demographic rates and functional traits over the 32 years of study (from 1992-2023; except for a few missing data in 2020 due to the covid-19 pandemic and unknown survival rates in 2023), using piecewise regressions. Using the '*segmented*' package (Muggeo 2008), we first assessed the optimal number of breakpoints according to lowest BIC by comparing models with 0 up to 6 breakpoints. Linear models (LM) or Generalized Linear Models (GLM) with binomial distributions (for survival rates and breeding proportion) were then performed over each defined segment or the entire period when no breakpoint was identified.

*Population density dependence and vegetation effects*

In order to differentiate the effects of forage availability and population density on population vital rates and functional traits, we ran 11 separate models each with a trait as dependent variable and population density, all three vegetation indices based on NDVI (mean NDVI over the season, NDVI relative growth rate [slope] and NDVI phenology) as independent variables. Mean NDVI over the season and NDVI growth slope were slightly correlated (Supplementary Data SD1: Fig. S4), but variance inflation factors remained low, so that all 4 independent variables were kept in the full models. We used either linear models (for emergence dates and masses, and female mass gain and litter size at weaning) or generalized linear models with a binomial distribution for binary variables (all survival rates and breeding proportion). Although we report statistical tests at a probability error threshold of 5%, results are discussed with regards to effect sizes rather than threshold p-values, which is more meaningful and now recommended by several studies (Halsey et al 2015; Nakagawa and Cuthill 2007). This was especially true because the relatively high number of models increased the risk of false discovery through multiple testing and because our main objective was to compare the relative importance of population density dependence and vegetation effects. To facilitate comparison of effect sizes, all explanative variables were standardized prior to analyses.

*Common trend analyses*

We used dynamic factor analyses (DFA) ('MARSS' function of the 'MARSS' package in R) to examine common trends of population functional traits and demographic rates and their relationships with population density and vegetation (Zuur et al. 2003). DFA is akin to PCA for time series data. It searches for common trends among multivariate time series and allows testing for important covariates which might influence the time-series. It includes two models:

an observation model and a process model. The process model represents the temporal common trends and is usually written as: $x_t = x_{t-1} + e_t$, where $e_t \sim MVN(0, Q)$. The observation model relates our temporal data $y_t$ to the common trends and can include covariates: $y_t = Zx_t + Dd_t + v_t$, where $d_t$ represents a vector of covariates and $v_t \sim MVN(0, R)$. Because of the seasonal biology of ground squirrels (cycles of hibernation and active seasons), we separated our analyses into **two different models**: *Model 1* considered the effect of the **preceding** year (covariates: population density$_{t-1}$ and vegetation indices $_{t-1}$) on hibernation emergence date and mass (the potentially associated variables). *Model 2* considered the effect of the **current** year (covariates: population density$_t$ and vegetation variables$_t$) on the other demographic rates and functional traits (potentially associated variables: litter size at weaning, mean juvenile survival, proportion of reproductive females in the population, mean female mass gain over the active season, mean adult male, adult female, and yearling female survival). To compare effect sizes, all independent variables were standardized prior to analyses.

For models 1 and 2, we ran a model selection procedure based on Akaike's information criterion corrected for small sample sizes (AICc). Candidate models were built based on their error matrix structures (diagonal and equal, diagonal and unequal, equal variance covariance, and unconstrained, see details in Supplementary Data SD1: Fig S5), the inclusion or not of the different covariates (from no covariates to all four: population density, mean NDVI, NDVI slope, NDVI phenology), and the number of common trends among potentially associated variables (from 1 to k-1, k being the number of examined variables). The model with the lowest AICc was considered the best fit. We also presented competing models that were similar in terms of strength of evidence, i.e. with a ΔAICc < 2 and AIC weights > 0.1 (W$_i$, corresponding to 10%, Wagenmakers and Farrell 2004; Burnham et al. 2011).

Finally, we explored whether the common trends found in the two DFAs could explain population growth from one year to the next, using linear models.

## RESULTS

### *Population density*

Over the 32-year long-term monitoring, overall mean spring population density (= population size on the constant 2.6 ha) averaged 65 ± 4 animals, ranging from 32 (1992 and 1993) to 120 (2002) individuals. Population density exhibited a strong increase from 1993-2002 followed by a sudden drop between the active period of 2002 and the active period of 2003 (from 120 to 37 individuals, GAM; edf = 6.0, F = 3.988, $p$ = 0.005, $R^2$ = 0.46, deviance explained = 56.4%, n = 32, Fig 1). After 2003, population size varied more moderately, generally recovering slowly from 33 individuals in 2004 to 96 individuals in 2022. On average, the population was composed of 61% ± 1% females (ranging from 44% to 81%). Over the 32 years of monitoring, there were between 14 and 77 females and between 7 and 43 males in the population.

### *Yearly vegetation growth and biomass*

Overall, the mean NDVI averaged from weekly values during the animal's active period (from the first emergences: week 14, to the approximate ending of the period of immergence: week 34) was 0.221 ± 0.008, ranging from 0.123 (2021) to 0.313 (2023) per year (n = 33 years). Years of high NDVI reflected "greener" years of overall higher productivity and biomass (Pettorelli et al. 2005), such as 2001, 2003, 2006, 2007, 2015, 2022 and 2023 (Fig 2.A). Contrarily, two periods displayed especially low NDVI during the active period of animals: 1991-1995 and 2020-2021. Vegetation growth rate (NDVI slope; Fig. 2B) was also variable over the years, averaging a slope of 0.019 ± 0.001, and ranging from 0.008 (1992) to 0.032 (2002 & 2011). The index used to reflect vegetation phenology varied greatly, from week 3 (Mid-January in 2016) to week 28 (early July) over the course of the study, overall averaging at week 19 (early May) ± 1 week. Vegetation green-up started particularly early in 2004, 2015 and 2016, but much later in 1992, 1998, 2010-2012 and 2022 (Fig 2.C).

*Temporal changes in population demographic rates and functional traits*

Demographic rates and functional traits exhibited marked interannual variations but also some long-term trends over the 32 years of study (Fig. 3; see Table 1 with segmented regression results). Indeed, ground squirrels emerged lighter and lighter over the course of the study (linear decrease from regression means over 32 years (-91g, -18% for males and -42g, -10% for females). This decrease coincided with later emergence dates in females (by 6.4 days over the 32 years) and in males (by 5.3 days though this delay in males was not significant, $P = 0.069$). Female adult survival and pup survival strongly decreased until 2002 before slowly recovering afterwards. In contrast, female mass gain during the reproductive season (*i.e.* from emergence of hibernation to weaning) followed a reverse trend, first almost doubling from 1992 to 2003 (+ 60g, range: 65 to 125g) before remaining stable for the rest of the study. No temporal trends were observed in the other 4 variables (litter size at weaning, breeding proportion, male adult survival and female yearling survival). However, these variables displayed very strong interannual variability (e.g., from 0 to 100% for yearling female survival).

*Population density dependence and vegetation effects*

In general, population density negatively impacted all vital rates and functional traits (except for female emergence date) although effects were only significant for female adult survival, litter size at weaning, breeding proportion and pup survival (Table 2; Fig. 4). Vegetation effects were contrasting and depended on the trait considered (Table 2; Fig. 4). The average vegetation biomass (as indicated by the mean NDVI over the reproductive season) negatively influenced litter size at weaning and pup survival, but had a positive effect on female yearling survival and little effect on other variables. Vegetation growth speed negatively affected all 4 survival rates, but had a positive effect on female mass gain. Finally, vegetation phenology had a positive effect on pup survival as well as female yearling and adult survival. Notably, pup survival was consistently influenced by density and NDVI (Fig. 4). The carry-over effects of population density and vegetation on the emergence of hibernation (date and mass) the next season all had high variances, resulting in no significant effects (bottom row of Fig. 4).

Finally, comparing effect sizes showed that no single parameter (population density or vegetation indices) could be seen as a primary influence on population vital rates and functional traits (Fig. 4). Indeed, while population density had a ubiquitous negative effect on vital rates and functional traits, it rarely displayed the strongest effect (only on litter size at weaning and breeding proportion). By contrast, the average NDVI was the variable with the strongest effect for female emergence mass and date. Vegetation growth slope had the strongest effect on female mass gain, female yearling survival, male adult survival and male emergence date and mass. Finally, vegetation phenology had the strongest effect on pup survival, female adult survival and female yearling survival (with late vegetative phenology favoring survival).

*Common trend analyses*

The first DFA model (*Model 1*) considered the effects of the previous year (population density and vegetation variables) on four functional traits, hibernation emergence date and mass of both males and females (Fig. 5, top). Model selection revealed the existence of one common trend in the 4 time-series, but no influence of environmental covariates (population density and vegetation indices) from the preceding year (*Model 1.1;* Table 3 and Supplementary Data SD1: Fig S6 for the fit of the data). The common trend reflected a negative association of emergence timing from hibernation and initial body mass, within both males and females (|factor loadings| > 0.2, Fig 5), showing that the sexes responded in a similar way but that body mass and date displayed opposite temporal trends. Overall, this trend showed a strong increase along the study period, *i.e.* a delay in spring emergence and a decrease in emergence mass, although there was an important drop in 2014-2015 (Fig 5). Finally, no effect of the common trend could be detected on year-to-year population growth (-0.04 ± 0.03, P = 0.187, $R_{adj}^2$=0.03).

The second DFA model considered the effects of the current year (population density and vegetation variables) on seven demographic rates and functional traits (Fig. 5, bottom). Model selection retained three competing models (ΔAIC < 2, *Models 2.1* to *2.3;* Table 3), all with the same matrix structure. However, the models differed in the number of common trends found between the seven traits considered. Model 2.1 and 2.3 displayed one common trend, while model 2.2 had two (see Fig. 5 & Supplementary Data SD1: Fig. S7 & S8). The common trend obtained with the best model correlated positively with trends for litter size at weaning, the proportion of reproductive females, pup survival, adult male and female survival and yearling survival. It correlated negatively with the trend for female mass gain between emergence and weaning (|factor loadings| > 0.2, Fig 5). The primary trend (models 2.1 & 2.3) was generally stable through the study period, but with a lot of short-term fluctuations (Fig 5). The reconstruction of patterns in the data using these models (along with their 95% confidence intervals) appeared to better fit raw data with models 2.1 and 2.3 than model 2.2 (Supplementary Data SD1: Fig S9). While the first two models both retained the mean NDVI as a covariate, the third had no environmental covariate. Notably, mean NDVI over the season negatively affected pup survival and litter size at weaning. Thus, in years when vegetation biomass was important, females surprisingly produced fewer offspring that survived less well to the next spring. Finally, the common trend positively affected population growth (0.24 ± 0.05, P < 0.001, $R_{adj}^2$=0.39) indicating that an increase in reproductive and survival rates (litter size at weaning, the proportion of reproductive females, pup survival, adult male and female survival and yearling survival) but a decrease in female mass gain led to an increase in population growth.

## DISCUSSION

We examined the effects of forage availability and population density on the demographic rates (yearling and adult survival, breeding success) and functional traits (phenology, body condition) of Columbian ground squirrels over a 32-year period. Our results showed a general negative effect of population density on vital rates. At the same time, vegetation also had major effects on demographic rates, though the effects varied and were sometimes in opposite directions, depending on individual sex or age. Further, we did not find evidence for carry-over effects of density and vegetation on adult functional traits (emergence phenology and body condition), but positive influences of vegetation phenology on offspring survival to the subsequent year. Finally, while temporal trends differed between traits, one common trend evidenced a positive covariation of all 6 vital rates (survival and reproduction) opposed to female mass gain over the reproductive season. This common trend was associated with the mean vegetation index (likely reflecting vegetation biomass) and in turn significantly explained variation in population growth.

The population started at relatively low density, followed by rapid growth. During this period, as density increased, initially high survival of adult females and their offspring declined, possibly reflecting increasing competition. This is similar to the experimentally induced density-dependent effects observed on reproduction and survival in Arctic ground squirrels, *Urocitellus parryii* (Karels and Boonstra 2000). Following an abrupt decline, population size increased more gradually, associated with more moderate changes in survival of females and their offspring. However, there appeared to be consistent temporal patterns that were not associated with the dramatic changes in population density. For instance, a delay in the timing of spring emergence was concurrent with a decline in body mass of adults, though only significant for females. This was also evident in the common trend analysis, which was not associated with any of the NDVI variables or population density from the previous year. The lack of carry-over effects suggests that the pattern of later spring emergences mostly reflected yearly spring conditions and a general delay of spring snow melt-off associated with cooler temperatures (Lane et al. 2012; Dobson 2016; Tamian et al. 2022; Thompson et al. 2023), rather than an influence of vegetation conditions from the previous year.

*Changes in demographic rates and functional traits*

Population density varied markedly over the course of our study, and exhibited a strong initial temporal increase followed by an abrupt fall and a subsequent more gradual growth and recovery. The population crash that occurred in 2002-2003 resulted from a combination of low survival rates and poor reproductive success (low breeding proportion and litter size at weaning), mostly due to overwinter predation by badgers on hibernating ground squirrels (167 badger digs recorded in spring 2003, an unusual event compared to other years; FSD, *unpublished data*). Yet, only survival was especially low survival during 2002-2003, especially for females and offspring. Reproduction and functional traits remained within the range of the rest of the study years, suggesting that these latter traits were not specific outliers in the observed trends of our analyses.

Given the marked changes in population density over the study years, we expected high variation in vital rates. Yet, we found little evidence for long-term changes in demographic rates, except for pups and adult females, for whom survival decreased in the first period of the study before increasing slightly. However, interannual variability in demographic rates was strong, especially regarding adult male survival, yearling female survival, and litter size. For instance, female yearling survival varied from 0 to 100% over the course of the study. By contrast, adult female survival exhibited relatively low interannual variation (see also Viblanc et al. 2022), as would be expected from long-lived species where adult survival may be buffered by life-history trade-offs against temporal variation (*viz.*, the "environmental canalization hypothesis," Wagner et al 1997; Gaillard et al. 1998; Gaillard and Yoccoz 2003). Such buffering might not apply to males, since they have highly variable reproductive success (produce up to 30 pups in one season, Raveh et al. 2010) and a shorter reproductive lifespan.

Next, we investigated the temporal variation of functional traits, which are typically highly variable and responsive to environmental changes (Jenouvrier et al. 2018). We found a delay in female emergence date and a decrease in emergence mass of males and females. This delay is counter to the pattern of earlier emergence by yellow-bellied marmots, *Marmota flaviventris*, where demographic rates were generally improved by a trend towards earlier emergence from hibernation (Ozgul et al. 2010). This means that individual Columbian ground squirrels emerged later but lighter in some years, but earlier and heavier in others. In years when emergence from hibernation was delayed, possibly due to unfavorable microclimatic conditions (Tamian et al. 2022), the inactive period spent fasting below ground would have lengthened, leading individuals to have fewer body reserves when they first came out of hibernation. Additionally, female ground squirrels emerged later and lighter from hibernation in recent years, showing changes in functional traits over time, suggesting plastic or micro-evolutionary responses to changes in the environment that are common to several species (e.g., reviews by Boutin and Lane 2014; Charmantier and Gienapp 2014; Schlichting and Wund 2014).

We also found strong interannual variation in female mass gain (somatic investment) over the breeding season, between emergence and weaning of the offspring. Since mass gain varied between means of less than 50g to more than 150g, but the length of this period varied little (from 53 to 57 days on average; Supplementary Data SD1: Fig S10), the strong and phenotypically plastic interannual variation that we observed was more likely related to changes in food availability/quality/phenology or to changes in energy allocation between soma and reproduction than to changes in amount of time needed to accumulate body mass (Dobson et al. 2023).

*Density-dependence and access to food resources*

In years when population density was high, mean litter size, the proportion of females reproducing, and the survival of all individuals decreased, at least to some degree. Increased population density might translate into increasing resource competition between individuals leading to either a decreased access to resources or higher aggression rates, stress and energy expenditure, and ultimately decreased survival and reproduction, as shown in other Columbian ground squirrel populations (Boag and Murie 1981; Dobson 1988, 1995; Dobson and Oli 2001; Oli and Dobson 2003) and other species (e.g., song sparrows, *Melospiza melodia*, Arcese and Smith 1988; Arctic ground squirrels, Karels and Boonstra 2000; osprey, Pandion haliaetus, Bretagnolle et al. 2008; Idaho ground squirrel, *Urocitellus brunneus*, Allison and Conway 2022). Similarly, the reproductive success of reindeer (*Rangifer tarandus*) has been shown to respond both to population density and vegetation green-up (Tveraa et al. 2013). Finally, we did not find density dependence for functional traits, such as emergence phenology and mass, and female mass gain between emergence and weaning. This could reflect a trade-off when competition is high, where females invest less in reproduction (Rubach et al. 2016), and shift their allocation strategy to invest more into their soma and maintain their mass gain through the season (e.g., by modulating offspring sex ratio; Barra et al. 2021; Kanaziz et al. 2022).

*Differential effects of vegetation on survival between sexes and reproductive status*

Vegetation indices had no significant direct effect on reproduction, neither the proportion of females weaning a litter, nor litter size at weaning. This might appear surprising, as Columbian ground squirrels are thought to be mostly income breeders (Broussard et al. 2005) that rely heavily on environmental resources during their short (3-4-month) breeding and subsequent active season. Previous experimental food-supplementation and empirical studies also indicated the importance of food resources on Columbian ground-squirrel life-histories (Dobson and Kjelgaard 1985a, 1985b; Dobson et al. 1986; Dobson 1988, 1995; Ritchie 1990). The amount of forage available to individuals, however, may have responded more strongly to population density and associated competition for food resources than to abundance of vegetation per se. The ground squirrels are highly phenotypically plastic in their demographic and functional traits (Dobson and Murie 1987; Dobson 1992; Dobson et al. 2023), and thus respond rapidly to changes in environmental conditions that include number of conspecific competitors, habitat changes, and changes in climatic conditions (Lane et al. 2012, 2019; Dobson et al. 2016; Thompson et al. 2023).

All three vegetation indices had important effects on ground squirrel survival rates. Surprisingly, rapid green-up of vegetation, reflected by the slope of NDVI in the spring/summer was associated with decreased survival of all age and sex groups. One possible hypothesis to this novel result may be that rapid vegetation growth leads to impoverished vegetation quality over the season, perhaps due to a rapid change in vegetation stages. Fast-growing vegetation might rapidly become less nutritive or digestible (higher contents of lignin for instance) at a critical time when ground squirrels need to fatten-up before hibernation.

In contrast, we found that later vegetation green-up was favorable to the survival of adult females, female yearlings, and young of the year. This would make sense if later vegetation growth allowed females to better support the high energy costs of lactation (Festa-Bianchet and Boag 1982; Kenagy and Barnes 1988; King et al. 1991; Speakman 2008), and allowed young of the year better late season growth and fattening conditions before their first hibernation, with beneficial survival outcomes (Murie and Boag 1984; Neuhaus 2000). Late timing of forage abundance appeared beneficial for offspring survival, while more rapid green up and greater overall forage abundance were detrimental. Non-reproductive yearlings also emerged from hibernation some 10 days later than older ground squirrels, on average (Tamian et al. 2022), so that a later green-up of the vegetation might also better match their energy requirements. Males emerged from hibernation first (Murie and Harris 1982; Tamian et al. 2022) and about 6 days before breeding females (Thompson et al. 2023). Males establish territorial boundaries and compete for mating opportunities shortly after emergence (Murie and Harris 1978; Manno and Dobson 2008; Raveh et al. 2010). Their energy requirements should thus be maintained throughout the active season, first for mating and then for fattening-up, so they should be less affected by vegetation phenology.

In conclusion, we found that density-dependence and characteristics of vegetation, remotely assessed through NDVI, had significant impacts on the demography and functional traits of Columbian ground squirrels. Overall forage abundance, measured in this way, also had a significant association with the rate of population growth. Increased density, however, had clear dampening influences on the population through reductions in reproduction and survival. Influences of vegetation were more nuanced and mostly focused on survival rates. Examining the relative contributions of density dependence and vegetation revealed no single influence of either parameter affecting population demography. Rather, effects and their relative importance varied according to the trait under study. Rate of vegetative growth and vegetative biomass, surprisingly, had negative effects on the survival of young through their first hibernation. The influence of later vegetative timing during the year was a more positive influence on survival of all ground squirrels. Common trend analysis, also surprisingly, showed a similar pattern with restricted reproduction and poor survival of the young during years when vegetation biomass should have been greatest. Thus, later timing of vegetation appeared to be a greater boon to the ground squirrels than increases in vegetative biomass.

## ACKNOWLEDGEMENTS

We are grateful to JO Murie for initiating the ground squirrel long-term study, for his critical comments on the study, and for his precious advice and council throughout the years. We are grateful to the Biogeoscience Institute, University of Calgary (E.A. Johnson and S. Vamosi, Directors; Judy Mappin-Buchannan and Adrienne Cunnings, Managers of the Kananaskis Field Stations; Kathreen Ruckstuhl, faculty member responsible for the R.B. Miller Field Station) for housing and facilities during fieldwork. We are especially grateful to E.A. Johnson for his continued support throughout the years. We are grateful to our colleagues and friends P. Neuhaus and K. Ruckstuhl for their critical insights during fieldwork, and to P. Neuhaus for his expert advice on ground squirrel ecology. The fieldwork was aided by many volunteers and students over the years, and we thank them for their excellent efforts. This research was supported by a USA National Science Foundation grant (DEB-0089473) and a fellowship grant from the Institute of Advanced Studies of the University of Strasbourg to F.S.D., a Natural Sciences and Engineering Research Council of Canada grant to J.O. Murie, a CNRS Projet International de Coopération Scientifique grant (PICS-07143), an AXA Postdoctoral Research

Fellowship, and a Fyssen Foundation research grant to V.A.V., and an Initiative d'Excellence (IDEX attractivité) research grant to C. Saraux. F.S.D. thanks the Région Grand Est and the Eurométropole de Strasbourg for the award of a Gutenberg Excellence Chair during the time of writing. A.T. was supported by a PhD scholarship from the Ministère de l'Enseignement Supérieur, de la Recherche et de l'Innovation. This study is part of the long-term Studies in Ecology and Evolution (SEE-Life) program of the CNRS

## SUPPLEMENTARY DATA

**Supplementary Data SD1.** Validation of the use of NDVI on our study site, details on the methodology applied to calculate NDVI indices, presentation of DFA matrices, fit of DFA trends on data, and results from competing DFA models (i.e. $\Delta AIC \leq 2$ & $w_i \geq 0.1$). https://doi.org/10.1007/s00442-024-05583-2.

## CONFLICT OF INTEREST

The authors declare no conflict of interest.

**Table 1. Statistical results of temporal changes in population demographic rates and functional traits.** Segmented regression analyses as a function of year were performed and the best number of breakpoints chosen according to BIC. Note that all survival rates as well as the breeding proportion were examined using binomial GLMs, while other traits were studied through LMs. Models in which the independent variable year was statistically significant at the $p \leq 0.05$ level appear in bold. The test value reported here is either a t-value (LM) or a z-value (GLM binomial). The last columns correspond to the adjusted r-squared or the percentage of deviance explained depending on whether a LM or GLM was performed, and the number of observations (years). When a breakpoint was present, estimates, along with test values, p-values and $R^2$ or explained deviance are presented for each regression.

|  | Variable | # breakpoints | Estimate ± SE | Test-value | P-value | $R^2$ | n |
|---|---|---|---|---|---|---|---|
| LM (t-value, $R^2$) | Litter size @ weaning | 0 | -0.02 ± 0.01 | -1.61 | 0.119 | 0.05 | 32 |
|  | **♀ Emergence Mass** | **0** | **-1.33 ± 0.39** | **-3.43** | **0.002** | **0.26** | **31** |
|  | **♂ Emergence Mass** | **0** | **-2.84 ± 0.67** | **-4.237** | **0.000** | **0.36** | **31** |
|  | **♀ Mass Gain** | **1 (2003)** | **4.99 ± 1.85** | **2.70** | **0.022** | **0.36** | **12** |
|  |  |  | -0.99 ± 0.89 | -1.11 | 0.280 | 0.01 | 20 |
|  | **♀ Emergence date** | **0** | **0.20 ± 0.08** | **2.37** | **0.025** | **0.13** | **31** |
|  | ♂ Emergence date | 0 | 0.16 ± 0.09 | 1.856 | 0.074 | 0.08 | 31 |
| GLM (z-value, % of deviance explained) | **Pup survival** | **1 (2002)** | **-0.15 ± 0.03** | **-5.32** | **< 0.001** | **33.6%** | **11** |
|  |  |  | **0.03 ± 0.01** | **3.02** | **0.003** | **7.5%** | **21** |
|  | Breeding proportion | 0 | -0.01 ± 0.01 | -1.58 | 0.114 | 2.4% | 32 |
|  | **♀ Adult survival** | **1 (2002)** | **-0.27 ± 0.06** | **-4.68** | **< 0.001** | **56.8%** | **11** |
|  |  |  | **0.05 ± 0.01** | **3.80** | **0.000** | **27.9%** | **21** |
|  | ♂ Adult survival | 0 | -0.01 ± 0.01 | -0.95 | 0.343 | 1.0% | 31 |
|  | ♀ Yearling survival | 0 | -0.01 ± 0.01 | -0.88 | 0.379 | 0.9% | 31 |

**Table 2. Effects of population density and vegetation (phenology, growth speed and average biomass) on Columbian ground squirrel's demographic rates and functional traits.** Estimates ± SE are reported along with t or z statistics and P-values.

|  | Population density | Vegetation biomass | Vegetation growth speed | Vegetation phenology |
|---|---|---|---|---|
| **Litter size @ weaning** | -0.213 ± 0.089<br>t = -2.37<br>P = 0.025 | -0.201 ± 0.107<br>t = -1.87<br>P = 0.072 | -0.088 ± 0.112<br>t = -0.79<br>P = 0.438 | 0.076 ± 0.101<br>t = 0.76<br>P = 0.456 |
| **♀ Emergence Mass** | -6.447 ± 4.070<br>t = -1.58<br>P = 0.126 | -6.724 ± 5.112<br>t = -1.31<br>P = 0.200 | 3.270 ± 5.473<br>t = 0.60<br>P = 0.556 | 4.199 ± 4.497<br>t = 0.93<br>P = 0.359 |
| **♂ Emergence Mass** | -4.847 ± 8.543<br>t = -0.57<br>P = 0.576 | -4.255 ± 10.730<br>t = -0.40<br>P = 0.695 | -5.007 ± 11.489<br>t = -0.44<br>P = 0.667 | -0.322 ± 9.439<br>t = -0.03<br>P = 0.973 |
| **♀ Mass Gain** | -3.285 ± 4.783<br>t = -0.69<br>P = 0.498 | -1.109 ± 5.795<br>t = -0.19<br>P = 0.850 | 12.919 ± 5.985<br>t = 2.16<br>P = 0.040 | -8.397 ± 5.388<br>t = -1.56<br>P = 0.131 |
| **♀ Emergence date** | 0.554 ± 0.954<br>t = 0.58<br>P = 0.567 | 0.552 ± 1.198<br>t = 0.46<br>P = 0.649 | -0.537 ± 1.283<br>t = -0.42<br>P = 0.679 | 0.303 ± 1.054<br>t = 0.29<br>P = 0.776 |
| **♂ Emergence date** | -0.369 ± 0.969<br>t = -0.38<br>P = 0.707 | -0.371 ± 1.217<br>t = -0.30<br>P = 0.763 | 0.888 ± 1.303<br>t = 0.68<br>P = 0.501 | 0.052 ± 1.070<br>t = 0.05<br>P = 0.961 |
| **Pup survival** | -0.144 ± 0.052<br>Z = -2.77<br>P = 0.006 | -0.189 ± 0.070<br>Z = -2.70<br>P = 0.007 | -0.173 ± 0.074<br>Z = -2.35<br>P = 0.019 | 0.224 ± 0.065<br>Z = 3.43<br>P < 0.001 |
| **Breeding proportion** | -0.217 ± 0.075<br>Z = -2.89<br>P = 0.004 | -0.069 ± 0.089<br>Z = -0.77<br>P = 0.438 | -0.152 ± 0.098<br>Z = -1.54<br>P = 0.122 | 0.088 ± 0.095<br>Z = 0.92<br>P = 0.355 |
| **♀ Adult survival** | -0.213 ± 0.080<br>Z = -2.66<br>P = 0.008 | 0.103 ± 0.109<br>Z = 0.94<br>P = 0.349 | -0.235 ± 0.115<br>Z = -2.04<br>P = 0.041 | 0.308 ± 0.100<br>Z = 3.10<br>P = 0.002 |
| **♂ Adult survival** | -0.162 ± 0.120<br>Z = -1.35<br>P = 0.177 | 0.180 ± 0.148<br>Z = 1.21<br>P = 0.224 | -0.478 ± 0.168<br>Z = -2.85<br>P = 0.004 | -0.016 ± 0.149<br>Z = -0.11<br>P = 0.914 |
| **♀ Yearling survival** | -0.218 ± 0.135<br>Z = -1.62<br>P = 0.105 | 0.356 ± 0.148<br>Z = 2.40<br>P = 0.016 | -0.929 ± 0.177<br>Z = -5.25<br>P < 0.001 | 0.373 ± 0.155<br>Z = 2.40<br>P = 0.016 |

**Table 3. DFA model comparisons.** Based on the ΔAIC < 2, the best two models for *Model 1* (Models 1.1 and 1.2, effects of density and vegetation indices in the **preceding** year on hibernation emergence date and mass) and the best three models for *Model 2* (Models 2.1 to 2.3, effects of density and vegetation indices in the **current** year on litter size at weaning, mean juvenile survival, proportion of reproductive females in the population, mean female mass gain over the active season, mean adult male, adult female, and yearling female survival). The considered covariates were included, along with the structure of the matrix, the number of common trends found in the model, the log Likelihood, the number of parameters in the model (K), the AICc and ΔAICc (difference compared to the best model), and the AICc weight ($W_i$, probability of the model being the best model). Models with the lowest AICc are indicated in bold. '*mean NDVI*' corresponds to the estimated vegetation biomass over the active season of ground squirrels.

| Model | Covariates | Matrix structure | # trends | logLik | K | AICc | ΔAICc | $W_i$ |
|---|---|---|---|---|---|---|---|---|
| **1.1** | Ø | Unconstrained | 1 | -122.0 | 14 | 276.2 | 0 | 0.81 |
| **2.1** | mean NDVI | Diagonal and unequal | 1 | -256.3 | 21 | 559.3 | 0 | 0.39 |
| **2.2** | mean NDVI | Diagonal and unequal | 2 | -249.1 | 27 | 560.1 | 0.8 | 0.26 |
| **2.3** | Ø | Diagonal and unequal | 1 | -265.6 | 14 | 561.2 | 1.9 | 0.15 |

**FIGURE LEGENDS**

**Fig. 1. Population size over the 32 years of the study.** Number of individuals in the population reflecting the population size and therefore density (the total surface is not changing) over the years, along with the fitted line and 95% confidence interval from GAM.

**Fig. 2. Fluctuations in vegetation biomass and growth over a 33-year period 1991-2023.** (A) Yearly vegetation biomass was estimated from the mean weekly value ± standard error of NDVI (greenness) over the active season of the squirrels, (B) vegetation rate of growth was estimated from the slope NDVI increase in each year, and (C) phenology index was defined as the week at which NDVI started increasing in each year (see Methods).

**Fig. 3. Temporal trends of population demographic rates and functional traits.** Temporal trends of litter size at weaning, pup survival, breeding proportion (*i.e.* proportion of females that weaned a litter), female emergence mass, male emergence mass, female mass gain, female emergence date, male emergence date, adult female survival, adult male survival, and yearling female survival. Segmented regression analyses were performed and the best number of breakpoints chosen according to BIC. Blue dots and segments represent breakpoints along with their 95% confidence intervals. Significant regressions are shown using black lines along with their 95% confidence intervals in dotted lines. Note that all survival rates as well as the breeding proportion were examined using binomial GLMs, while other traits were studied through LMs. Colors correspond to variable colors in the DFA analysis.

**Fig. 4. Effect sizes of population density and vegetation indices (mean biomass, phenology and growth speed) on vital rates and functional traits.** Effect sizes correspond to linear estimates for linear models (litter size at weaning, female mass gain, male and female emergence dates, male and female emergence masses) and to odds ratio for all generalized linear models with binomial error distribution (all 4 survival rates as well as breeding proportion). They are represented along with their 95% CI. All four independent variables were standardized prior to models. Positive effects are indicated in blue and negative effects in red. *P < 0.05, ** P < 0.01, *** P < 0.001.

**Fig. 5. Dynamic factor analyses of Columbian ground squirrel vital rates and functional traits.** Higher panel: *Model 1.1* on emergences (male and female emergence dates and masses), model selection retained no environmental covariate; Lower panel: *Model 2.1* with the litter size, pup survival, reproduction proportion, female mass gain, adult female survival, adult male survival, and yearling female survival; model selection retained mean vegetation biomass as an environmental covariate. The left column displays the common trend, the middle one factor loadings of the traits on this common trend and the right column displays effect sizes with their 95% confidence interval of the environmental covariates retained in the model. Positive loadings correspond to positive relationship with the trend, and negative loadings to negative relationship with the trend (colored bars). Effect sizes correspond to estimates of the D matrix (see methods), i.e. here the effect of standardized mean NDVI on all 7 standardized traits for model 2.1, while Ø indicates the absence of environmental covariates in model 1.1.

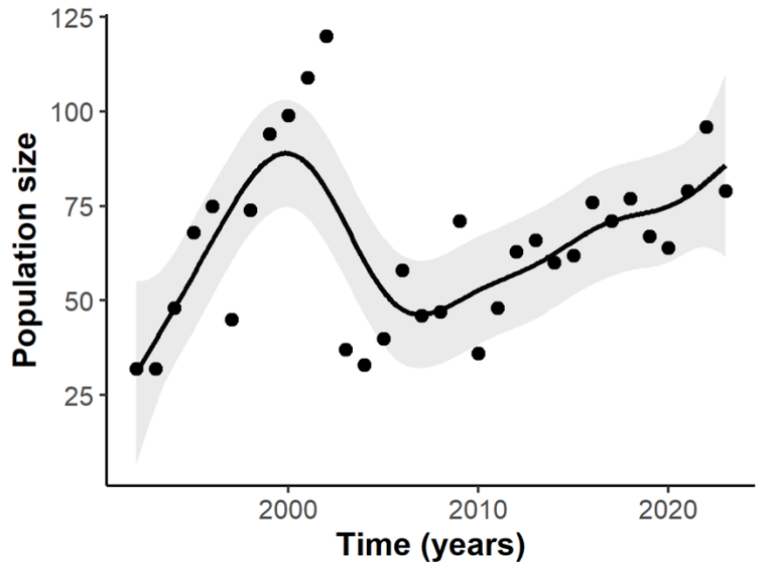

Fig 1.

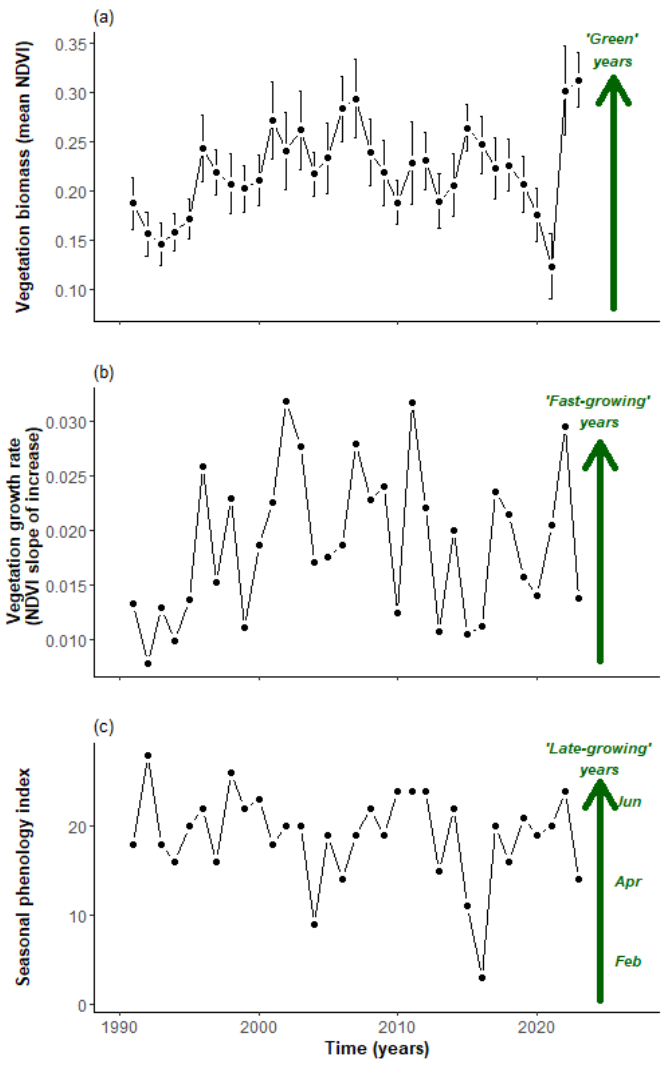

Fig 2.

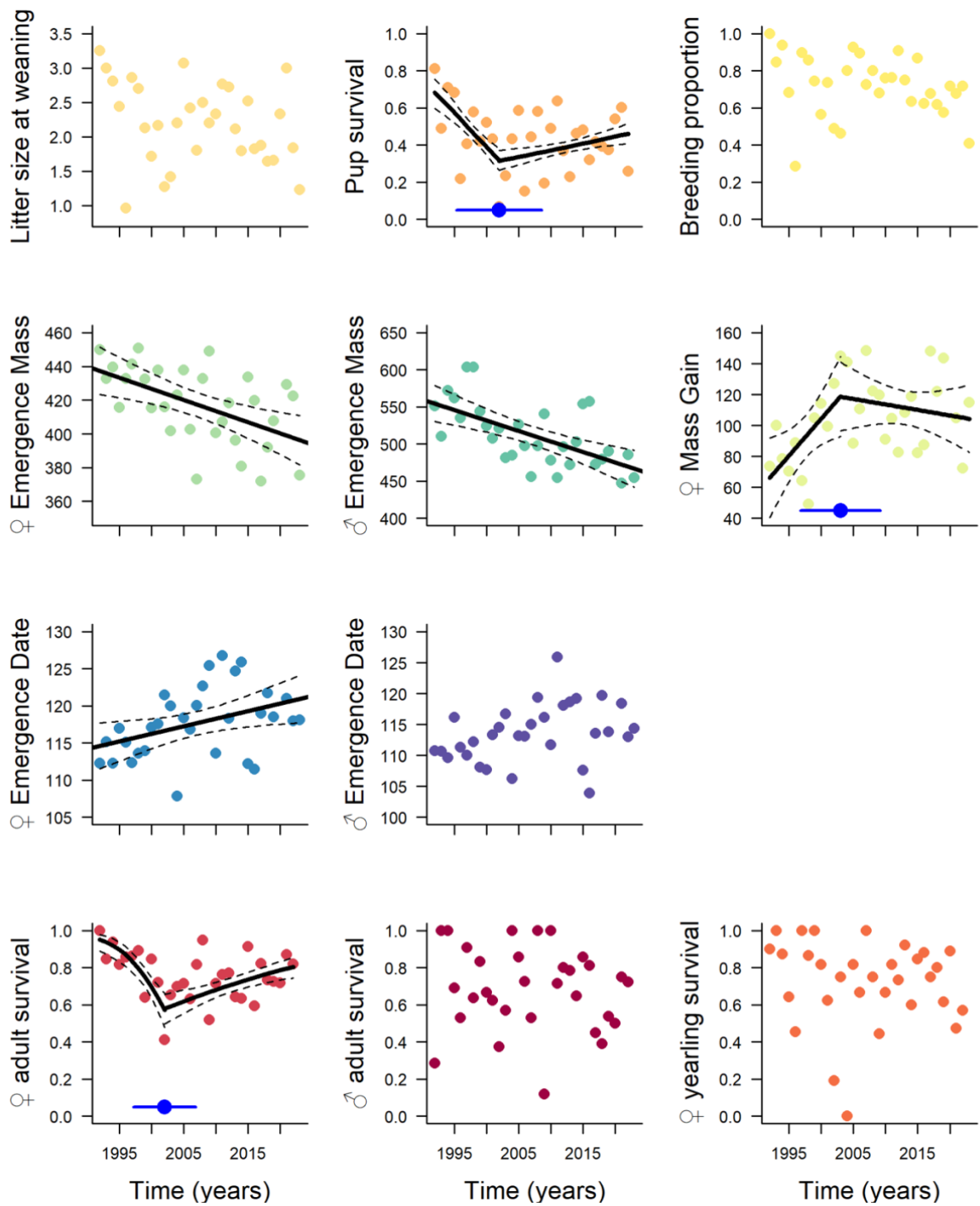

Fig 3.

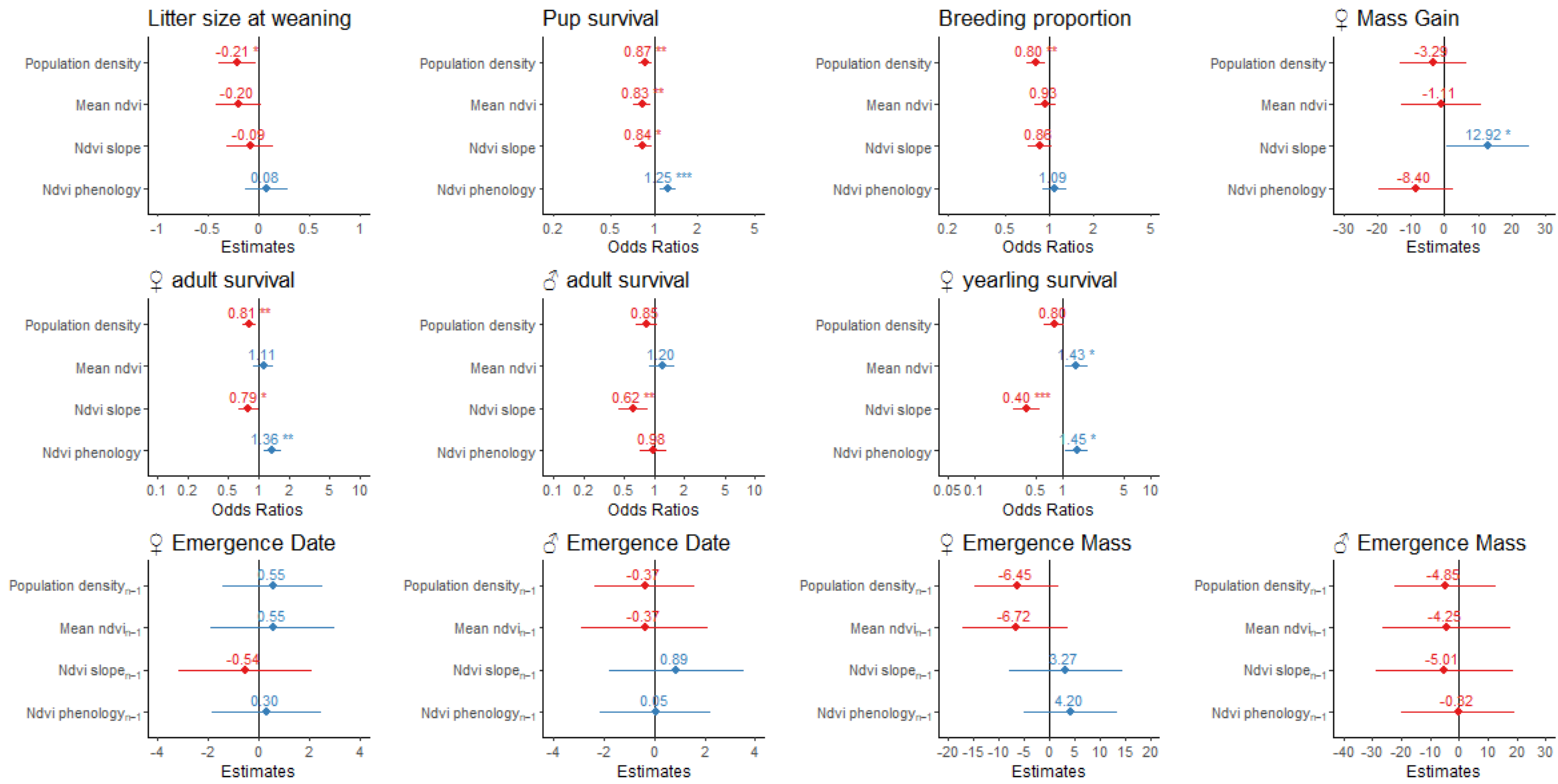

Fig 4.

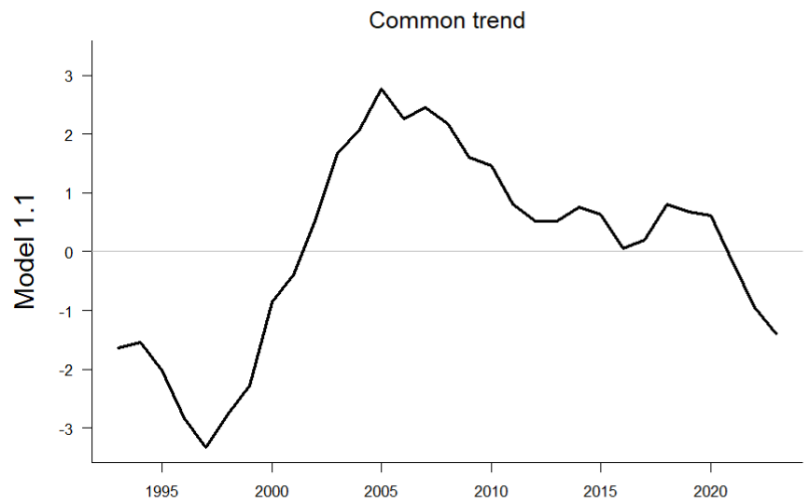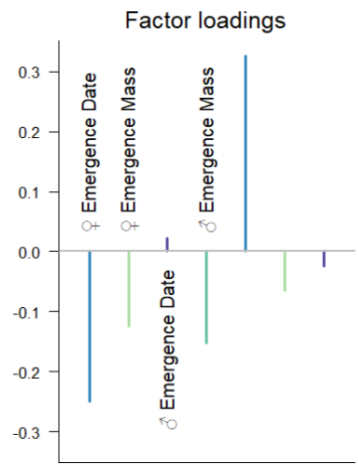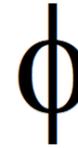
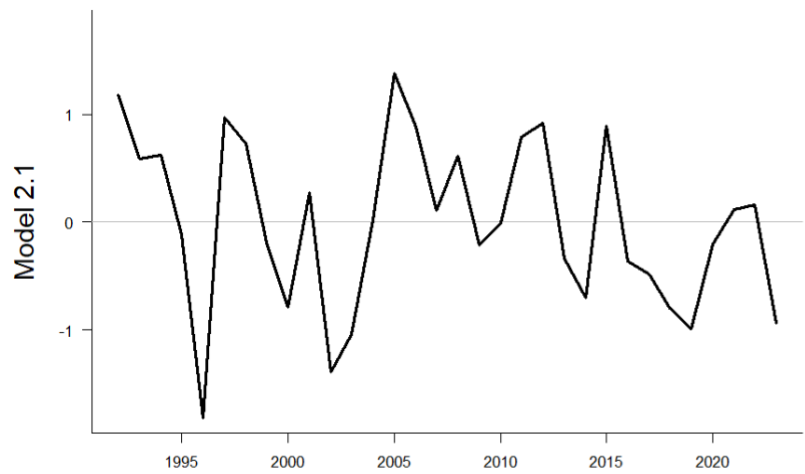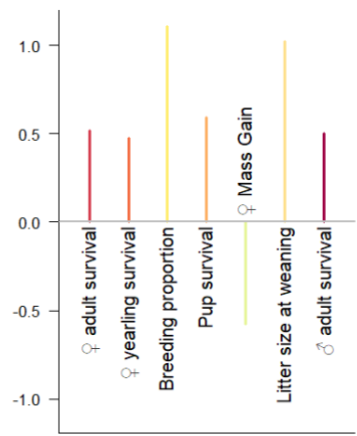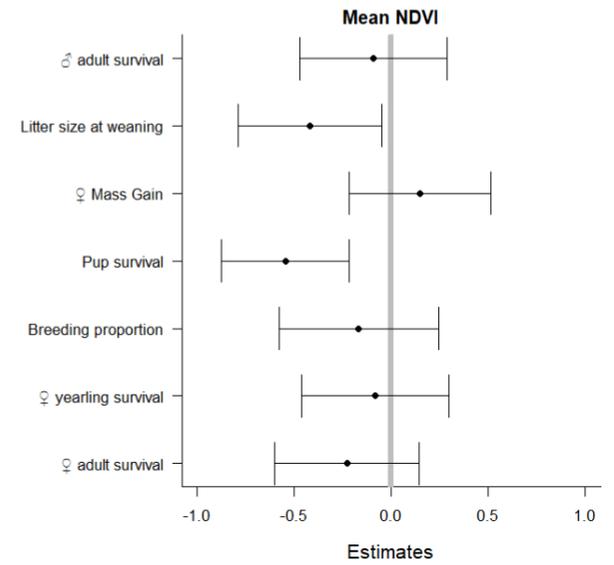

Fig 5.